# Stabilization of Co oxide during oxygen evolution in alkaline media by the introduction of Mn oxide

Javier Villalobos,[a] Dulce M. Morales,[a] Denis Antipin,[a] Götz Schuck,[b] Ronny Golnak,[c] Jie Xiao,[c] Marcel Risch[a]*

[a]  J. Villalobos, D.M. Morales, D. Antipin, M. Risch
     Nachwuchsgruppe Gestaltung des Sauerstoffentwicklungsmechanismus
     Helmholtz–Zentrum Berlin für Materialien und Energie GmbH
     Hahn-Meitner Platz 1, Berlin 14109, Germany
     E-mail: marcel.risch@helmholtz-berlin.de
[b]  G. Schuck
     Abteilung Struktur und Dynamik von Energiematerialien
     Helmholtz–Zentrum Berlin für Materialien und Energie GmbH
     Hahn-Meitner Platz 1, Berlin 14109, Germany
[c]  R. Golnak, J. Xiao
     Department of Highly Sensitive X-ray Spectroscopy
     Helmholtz–Zentrum Berlin für Materialien und Energie GmbH
     Albert-Einstein-Straße 15, Berlin 12489, Germany

Supporting information for this article is given via a link at the end of the document.

**Abstract:** Improving the stability of electrocatalysts for the oxygen evolution reaction (OER) through materials design has received less attention than improving their catalytic activity. We explored the effect of Mn addition to a cobalt oxide for stabilizing the catalyst by comparing Na-containing $CoO_x$ and $(Co_{0.7}Mn_{0.3})O_x$ films electrodeposited in alkaline solution. The obtained disordered films were classified as layered oxides using X-ray absorption spectroscopy (XAS). The $CoO_x$ films showed a constant decrease in the catalytic activity during cycling, confirmed by oxygen detection, while that of $(Co_{0.7}Mn_{0.3})O_x$ slightly increased as measured by electrochemical metrics. These trends were rationalized based on XAS analysis of the metal oxidation states, which were $Co^{2.8+}$ and $Mn^{3.7+}$ near the surface after cycling. Thus, adding Mn to $CoO_x$ successfully stabilized the catalyst material and its activity during OER cycling. The development of stabilization approaches is essential to extend the durability of OER catalysts.

## Introduction

The use of fluctuating renewable sources, such as sunlight and wind, limits renewable energy production due to the lack of efficient energy storage systems. A promising solution is chemical energy storage using hydrogen obtained by water spitting.[1,2] The most daunting challenges in the efficient use of water splitting are finding highly active electrocatalysts to overcome the slow kinetics of the oxygen evolution reaction (OER), exhibiting, in addition, sufficient stability under the harsh operating conditions.[3–5] In the last decades, most of the research in this field has been focused on developing new electrocatalysts or improving the catalytic properties of the already known electrocatalysts in terms of catalytic activity, which has been the primary parameter of interest.[6] Nevertheless, stability should not be considered a parameter of secondary importance since novel long-term stable catalysts are urgently needed for technical applications.

Alkaline electrolyzers are a mature technology for low-temperature electrolysis with a target stack lifetime of 25 years.[7] Many amorphous transition-metal oxides (ATMO) are thermodynamically stable in alkaline electrolytes and show high catalytic activity.[8–11] In academic research, ATMO based on earth-abundant metals own many advantages over the benchmark Ir- or Ru-based oxides, such as high catalytic activity, high stability and low-cost.[12–14] We define stability herein as the absence of catalyst corrosion,[15] erosion[16,17] and blockage of active sites (e.g. by oxygen bubbles),[18–24] for which a first indication is a lack of change in activity over time, e.g. measured by cyclic voltammetry.[25–27] Yet, the discussion of stability requires additional measurements to determine dissolved cations,[28,29] as well as changes in the catalyst composition,[20,30] morphology[31–33] and structure.[34,35]

Co-based ATMO have attracted particular attention due to their high catalytic activity. However, Co oxides suffer from poor electrical conductivity[36,37] and tend to corrode over time.[38] The introduction of a second transition metal into the Co-based oxides alters the electronic structure and potentially also modifies the atomic rearrangement, affecting catalysis and corrosion resistance.[39–42]

Introducing Mn as a second metal has enhanced stability of perovskite-like oxides[43] and electrodeposited mixed metal oxides,[44] which has been attributed to separation of the structural framework from the catalytically active site(s).[44] The activity was also enhanced by adding Mn in some reports, e.g., the introduction of 25 % of Mn into the $Co_3O_4$ spinel structure showed an overpotential decrease from 368 mV to 345 mV (at a current density, $j$, of 10 mA cm$^{-2}$).[39] Menezes and collaborators[41] compared the current stability of the spinels $CoMn_2O_4$ and $Co_2MnO_4$. The current of both catalysts remained mostly constant after 30000 s, yet $Co_2MnO_4$ (containing more Co than Mn) showed a higher catalytic current. The role of Mn in layered Co oxide has been attributed to the modulation of the electronic properties, resulting in a more efficient charge-transfer.[45] Sugawara et al.[46] attributed the differences in activity and stability among CoMn oxides to differences in the metal coordination in layered, tunnel and spinel oxides. In summary, the possible roles of Mn in the Co oxide structures are still insufficiently understood and they cannot be predicted *a priori* the extent to which the addition of Mn will beneficially affect activity, stability or both.

In this study, we extended our previously reported alkaline electrodeposition method[47] to Na-containing $CoO_x$ and $(Co_{0.7}Mn_{0.3})O_x$ films without long-range order. Activation and degradation process were studied in the films during cyclic voltammetry and open-circuit conditions in 0.1 M NaOH. During cycling, we observed a slight increase in the current for $(Co_{0.7}Mn_{0.3})O_x$, whereas $CoO_x$ showed a decrease in the catalytic current. The post-mortem samples were analyzed by XAS to understand the observed electrochemical changes. We conclude that the introduction of Mn oxide into the Co oxide structure increases the stability of the films, both structurally and catalytically.

## Results and Discussion

$CoO_x$ and $(Co_{0.7}Mn_{0.3})O_x$ films were deposited on glassy carbon (GC) rods following a previously reported protocol from our group for the electrodeposition of $MnO_x$ films in alkaline pH.[47] Like Mn and other metals in water-based solutions, Co may spontaneously deposit as oxides or hydroxides in alkaline media. Thus, tartrate ions are included in the electrodeposition electrolyte as a complexing agent to stabilize the metal ions within the electrodeposition procedure. Using the same ions ($Na^+$, $OH^-$) in the electrolyte for both the electrodeposition and the catalytic investigation prevents the plausible anionic exchange between the catalytic material and the electrolyte during OER.[48]

The galvanostatic electrodeposition of the films was carried out in a three-electrode cell using a commercial (unrotated) RDE holder (Figure 1a). A constant current of 0.15 mA cm$^{-2}$ was applied until a charge density of 40 mC cm$^{-2}$ was reached. $CoO_x$ reached a minimum steady-state potential of 1.45 V vs. RHE after about 20 s, whereas $(Co_{0.7}Mn_{0.3})O_x$ reached 1.91 V vs. RHE after the same time. The different potentials suggest the formation of different materials.

Since the glassy carbon (GC) rods used as substrates have a small surface area, the films were also deposited on larger graphite foil (GF) following the same protocol for further XAS and energy-dispersive X-ray spectroscopy (EDX) characterization. In both cases a steady current was reached after several seconds, yet the absolute potentials differ between both substrates (Figure S1), likely because the electrodeposition potential also depends on the substrate's properties, e.g., electrical conductivity. The steady-state was reached with a potential shift of about 0.4 V for $(Co_{0.7}Mn_{0.3})O_x$ and 0.1 V for $CoO_x$ to higher potential on GC relative to GF. Electrochemical experiments on both substrates were performed to exclude that these electrodeposition potential variations affect the catalytic properties of the films. These results are discussed below.

The films were characterized by scanning electronic microscopy (SEM) to check the coverage and homogeneity of the film on the substrate. The SEM images (Figure S2) showed a full coverage of the film over the GC surface. Moreover, EDX was used to map the homogeneous distribution of the two metals on the film-deposited graphite foil (Figure 1b). The average ratio of Co/Mn in $(Co_{0.7}Mn_{0.3})O_x$ was 2.45±0.06, which we estimated as an average of the composition observed in different regions of three samples. The Co/Mn ratio indicates that out of the total metal sites (Co+Mn), approximately 70±5 % correspond to Co and 30±5 % to Mn.

Moreover, the EDX spectrum showed high content of carbon (from the carbon-based substrate), oxygen (from the substrate and the film), and sodium (coming from the electrolyte).

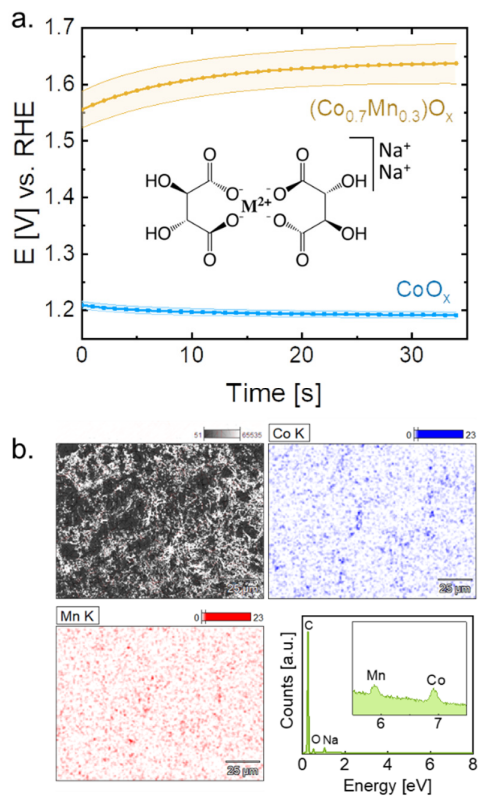

**Figure 1.** a. Electrodeposition chronoamperometry of $CoO_x$ and $(Co_{0.7}Mn_{0.3})O_x$ films on glassy carbon in NaOH 0.1 M until a charge of 40 mC cm$^{-2}$. The inset shows the coordination complex of divalent metal ($M^{2+}$) due to tartrate ions b. EDX map of the $(Co_{0.7}Mn_{0.3})O_x$ film: SEM image (top left), Co map (top right), Mn map (bottom left), and EDX spectrum (bottom right). Dataset in ref.[49]

No substantial morphology differences were observed between the pristine $CoO_x$ and $(Co_{0.7}Mn_{0.3})O_x$ films (Figure S2). Additionally, no significant morphological changes were observed in comparison with the previously reported $MnO_x$.[47] In summary, the protocol of electrodeposition in alkaline pH was successfully extended to the deposition of Na-containing $CoO_x$ and $(Co_{0.7}Mn_{0.3})O_x$ films without long-range order.

The catalytic stability of the films during OER catalysis was evaluated by cyclic voltammetry (CV) in a three-electrode cell in a rotating-ring disk electrode station (RRDE), comprising the $CoO_x$- and $(Co_{0.7}Mn_{0.3})O_x$-covered GC rod as the disk electrode and a Pt ring as the ring electrode (protocols are shown in Table S1 for GC and Table S2 for GF). The CV series of $CoO_x$ and $(Co_{0.7}Mn_{0.3})O_x$ (Figure 2a, 2b, S3) were collected in 0.1 M NaOH with a scan rate of 100 mV s$^{-1}$ for a total of 100 cycles. Similar scan rates and number of cycles are typical conditions for film activation during OER.[48,50–54] Meanwhile, the Pt ring was set at a constant potential of 0.4 V vs. RHE for oxygen reduction.[55] With this setup, the oxygen produced at the disk during continuous potential cycling is consequently reduced at the ring. Since the exponential increase in the ring current density ($j_{ring}$) due to reduction of oxygen matches that observed at the disk electrode ($j_{disk}$), the latter can be associated with oxygen evolution. At the same time, a rough estimation of the OER onset potential can be determined, which we defined at the potential where the ring current reaches 0.15 µA cm$^{-2}$ during the second cycle. For CoO$_x$, the OER onset is around 1.64±0.02 V vs. RHE, whereas for (Co$_{0.7}$Mn$_{0.3}$)O$_x$ is 1.66±0.01 V, an negligible difference within error. The overpotential *of the electrode* ($\eta_{10}$) was calculated at a specific current density per geometric area, $j$=10 mA cm$^{-2}$, which is chosen based on the current drawn by a solar-to-fuel device with a 10 % of efficiency under one sun illumination.[56] It is important to note that $\eta_{10}$ is a helpful metric to compare electrodes but it cannot be used to compare the intrinsic properties of different materials.[56,57] In CoO$_x$, $\eta_{10}$ was 466±15 mV after 2 cycles, and it increases to 520±19 mV after 100 cycles. In (Co$_{0.7}$Mn$_{0.3}$)O$_x$ $\eta_{10}$ was 510±30 mV after the first 2 cycles and 500±27 mV after 100 cycles, i.e. remained constant within error. Electrodes with similar composition (Co,Mn- and Co-based oxide) investigated in alkaline pH (13-14),[58–62] showed $\eta_{10}$ in a range of 320-430 mV, which tend to increase under OER conditions, thus agreeing with the observations herein. Although the introduction of a second transition metal into the Co oxide structure has reduced the overpotential in some cases,[39,63–65] while it was increased in other cases.[66]

Catalytic trends can also be followed using the maximum current density ($j_{max}$, at an ohmic drop-corrected potential (E-iR$_u$) of approximately 1.73 V vs. RHE) over cycling. In the case of CoO$_x$, the disk $j_{max}$ decreases over cycling; about -33±15 % of the initial current is lost after 100 cycles. This effect is also observed at the ring current, where the current drops about -35±14 % compared to the initial value, indicating that the drop in $j_{max}$ is (mainly) due to deactivation of the catalyst film during cyclic voltammetry. In contrast, $j_{max}$ of the (Co$_{0.7}$Mn$_{0.3}$)O$_x$ disk remained mostly stable (with slight increase) over 100 cycles compared to the initial value, about +10±1 % at the disk and +11±4 % at the ring, indicating a stabilization of the catalytic current during cycling voltammetry, namely, a higher amount of oxygen is produced and detected at the disk and ring electrode, respectively. The CV series were also collected with a wider potential range to confirm that a possible incomplete reduction does not affect the $j_{max}$ trends during cycling (Figure S4).

The CV experiment was extended after the 100 cycles with 30 minutes at open circuit potential (OCP) and 10 additional cycles in the same potential range. The goal of introducing an OCP break between the two series of CV is to identify if the catalytic current suffers changes after the OCP period, therefore distinguishing reversible and irreversible changes in the catalyst.[47,67] The current density at selected potentials was plotted as a function of number of cycles for a more detailed analysis of the trends (Figure 3). Note that both x and y axis are presented in a logarithmic scale. The capacitance was corrected by normalizing the average between the anodic and cathodic scans by the difference between the cathodic and anodic current at E-iR$_u$=1.5 V vs. RHE, $\Delta i_{1.5V}$, (Figure 3a,c).[57] This represents a rough approximation of the capacitance, which is more commonly estimated by a systematic experiment that involves recording CVs at several scan rates.[68] Yet, it allows tracking changes in the surface area with cycling. The current trend was analyzed at three different potential values, which were selected based on the estimation of the oxygen evolution onset: no OER (1.55 V vs. RHE), onset of OER (1.64 V for CoO$_x$ and 1.66 V vs. RHE for (Co$_{0.7}$Mn$_{0.3}$)O$_x$), and OER (1.70 V vs. RHE). The normalized current, $i/\Delta i_{1.5V}$, follows different exponents (slopes in the logarithmic plot) depending on the cycle number, the selected potential, and whether the cycles were recorded before or after the OCP break. Thus, a negative exponent represents a current decrease, an exponent close to zero represents stable current, and a positive exponent represents a current increase with cycling. Since the exponent depends on the cycle number, the 100 cycles before the OCP break were split into three regions, 1 (1-10$^{th}$ cycle), 2 (11-50$^{th}$ cycle) and 3 (51-100$^{th}$) for analysis.

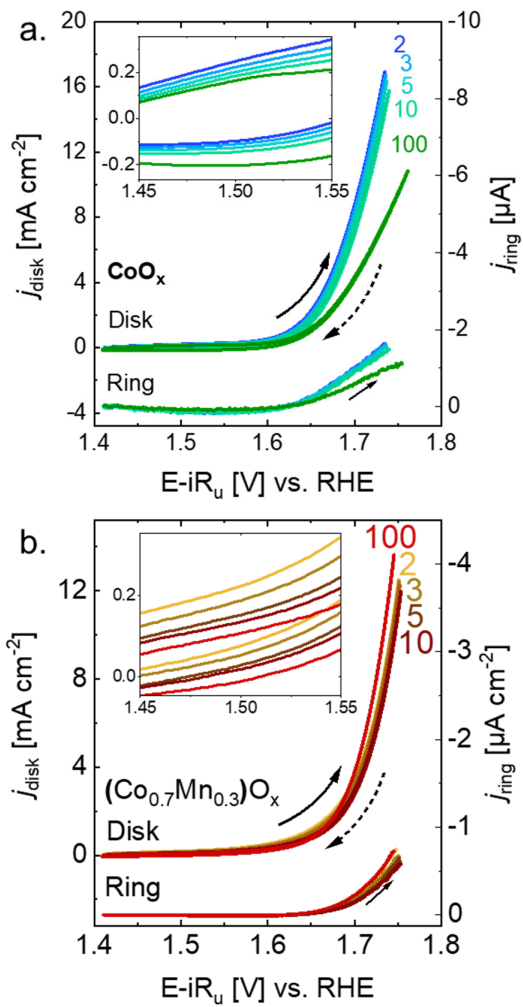

**Figure 2.** Series of CV performed on: a. CoO$_x$-covered disk and b. (Co$_{0.7}$Mn$_{0.3}$)O$_x$-covered disk. The data was collected with a scan rate of 100 mV s$^{-1}$ in 0.1 M NaOH with an electrode rotation 1600 rpm and a constant potential of 0.4 V vs. RHE at the Pt ring. Dataset in ref.[49]

The current trends of CoO$_x$ show a constant decrease during the first 100 cycles at each of the selected potentials (Figure 3a). After the OCP break, the current partially recovers at both $i_{1.55}/\Delta i_{1.5V}$ and $i_{1.64}/\Delta i_{1.5V}$, which can be observed by the position of the solid squares (after OCP) below the open squares (before OCP) in Figure 3a. Only at $i_{1.70}/\Delta i_{1.5V}$, the current fully recovers after the OCP break (Figure 3a). This recovery was also observed in the $j_{ring}$ (Figure S5). The decay exponents did not strongly vary among the three different regions (Table S3). Since the OER catalytic current is the major current component at $i_{1.70}/\Delta i_{1.5\,V}$, a current trend recovery could be observed. Whereas, at $i_{1.64}/\Delta i_{1.5\,V}$ and $i_{1.55}/\Delta i_{1.5\,V}$, there might be a significant current contribution from irreversible processes, e.g., Co redox changes, which cannot be recovered after the OCP break.

The current trends of $(Co_{0.7}Mn_{0.3})O_x$, show a clear difference in the exponent, depending on the selected potential. At the non-OER potential, $i_{1.55}/\Delta i_{1.5\,V}$ decreases over cycling in all three regions. Whereas, at the OER onset, $i_{1.66}/\Delta i_{1.5\,V}$ showed a negative exponent in region 1 and 2 and became positive in region 3. For $i_{1.70}/\Delta i_{1.5V}$, the exponent changed from a positive value close to zero in region 1 and kept increasing towards more positive values in region 2 and 3, indicating the stabilization of the current (with a slight activation) at this potential (as also observed with the ring and disk $j_{max}$ in Figure 2b). The exponent values are summarized in Table S3.

The trends during continuous potential cycling of the films on GF were also plotted (Figure S6) and showed trends similar to those observed on GC. Thereby, we confirmed that the variations in the electrodeposition potential due to different substrates did not significantly affect the current trends during cycling.

The Tafel slope ($b=\partial log i / \partial E$) indicates the scaling of kinetic currents with applied potential where a low value indicates that a large increase in current can be achieved with a small increment in overpotential (i.e., far from equilibrium). Its values can be rationalized based on mechanistic considerations such as the rate-limiting step and the populations of surface intermediates.[69,70] For instance, for the OER, a value of 60 mV dec$^{-1}$ is associated with a chemical rate-limiting step with an electrochemical pre-equilibrium. A value of 120 mV dec$^{-1}$ is related to an electrochemical rate-limiting step, and a value much greater than 120 mV dec$^{-1}$ is due to chemical limiting step or poor material conductivity.[71]

The Tafel plots were analyzed for both materials, $CoO_x$ and $(Co_{0.7}Mn_{0.3})O_x$, at the OER potential range (1.70-1.76 V vs. RHE). From the plots, Tafel slopes were determined and plotted as a function of the number of cycles (Figure 4). A representative calculation is shown in Figure S7 with averaged parameters shown in Table S4. The Tafel slope as function of potential was also plotted (Figure S8). Considering that a scan rate of 100 mV s$^{-1}$ may be too fast to establish a complete chemical equilibrium, the produced intermediates can be shifted towards the oxidized sites during the cathodic scans (since high potentials are applied) if an electrochemical step is part of the OER mechanism. Thus, only the anodic scans are used to estimate the Tafel slopes. At $CoO_x$, the Tafel slope was around 135±10 mV dec$^{-1}$ in the initial 10 cycles and it increased insignificantly to 158±25 mV dec$^{-1}$ at the 100$^{th}$ cycle. Yet, the slope went down to 133±11 mV s$^{-1}$ after the OCP break. At $(Co_{0.7}Mn_{0.3})O_x$, the Tafel slope is mostly constant during 100 cycles and 10 cycles after OCP break, with a value of 89±2 mV s$^{-1}$. Typical Tafel slope values for layered Co oxides are about 60 mV dec$^{-1}$,[33,48,72,73] whereas layered Mn oxides show values between 60 mV dec$^{-1}$ and 180 mV dec$^{-1}$.[47,74–76] Tafel slope values between 60 and 120 mV dec$^{-1}$ are not predicted by common kinetic modeling. However, variations in the material's charge transfer coefficient ($\alpha$) would lead to different Tafel slope values[70,77], as well as non-catalytic side reactions such as metal redox independent of catalysis,[69] and changes in coverage and/or electrical conductivity during the potential scan.[78]

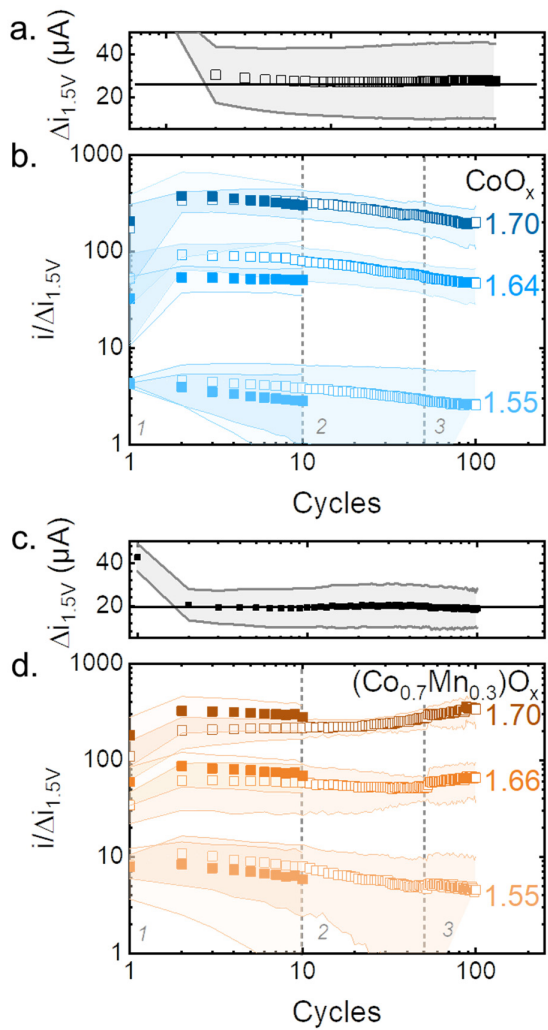

**Figure 3.** Average $\Delta i_{1.5\,V}$ of all samples as function of cycles for the first 100 cycles for a. $CoO_x$ and c. $(Co_{0.7}Mn_{0.3})O_x$. Average current ratio $i/\Delta i_{1.5\,V}$ of all samples as function of cycling at selected potentials for b. $CoO_x$ and d. $(Co_{0.7}Mn_{0.3})O_x$. The data was extracted from the first 100 cycles (open squares) and from 10 cycles recorded after 30 min of OCP break (solid squares). The light-colored areas represent the standard deviation of three samples. The dashed lines separate three regions: 1, 2 and 3. Dataset in ref.[49]

In summary, $CoO_x$ films deactivated slightly during 100 cycles, yet the current fully recovered at the catalytic potential (1.70 V vs. RHE) after a 30-minute OCP break, indicating reversible changes likely due to coverage changes, for instance, unreacted intermediates.[67] The Tafel slope remained larger than 120 mV dec$^{-1}$ and increased over cycling, suggesting a change in the coverage over time. In contrast, the current at OER potentials and the Tafel slope values of $(Co_{0.7}Mn_{0.3})O_x$ were mostly stable with cycling.

$CoO_x$ and $(Co_{0.7}Mn_{0.3})O_x$ were studied under the same conditions, yet they show different catalytic properties and current trends with cycling, due to the presence of Mn. Both metals, Mn and Co, are well known as OER catalysts, therefore it is likely the Mn (as well as Co) plays an important role in the catalytic process. The OER activity of Co and Mn has been reported for bimetallic oxides.[39,41,45,46] Thus, the study of both metal structural positions is necessary for a better understanding of the changes observed over cycling.

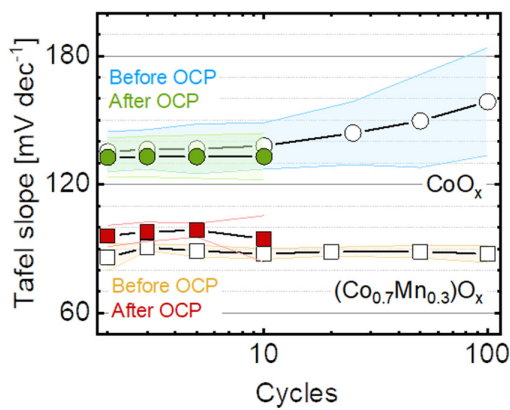

**Figure 4.** Averaged Tafel slope as function of cycle number before (open symbols) and after (solid symbols) a 30-minute break at OCP. The light-colored areas represent the standard deviation of three samples. The Tafel slopes were calculated in the potential range between 1.70 V and 1.76 V vs. RHE. Dataset in ref.[49]

XAS experiments were performed to investigate irreversible structural changes in the catalyst due to cyclic voltammetry. The absence of crystallinity in the films requires XAS experiments to analyze possible structural changes, which is not possible by diffraction-based techniques. The Co-K and Mn-K edge were used to study the bulk of the material since the radiation deeply penetrates the catalyst. Using XANES (X-ray absorption near edge structure), changes in the averaged oxidation state were identified and using the EXAFS (extended X-ray absorption fine structure), changes in the local structure were tracked. Moreover, the Co-$L_3$ and Mn-$L_3$ edge were used to study the oxidation state at the near-surface region by the electron yield.

The FT (Fourier transform) of EXAFS spectra collected on $CoO_x$ and $(Co_{0.7}Mn_{0.3})O_x$ showed typical features of layered hydroxides (Figure 5a, 5b) in both edges, Co-K and Mn-K. Two prominent peaks were identified: a M-O peak of around 1.87 Å, and a M-M peak of around 2.81 Å, where M is either Mn or Co. The phase functions were simulated using several reasonable structural models, such as birnessite ($MnOOH \cdot xH_2O$), heterogenite (CoOOH), and $Co(OH)_2$. Three relevant parameters were obtained from the simulations: $N$, which is related to the number of neighboring atoms around the absorber atom, $R$, related to the averaged interatomic distance between the absorber atom and the scatter, and $\sigma$ (Debye-Waller factor), associated with the distance distribution in a disordered material. The simulation parameters are summarized in Table 1 and Table 2, and the corresponding k-space spectra are shown in Figure S9. Note that that reduced distance is shorter than the precise distance obtained by EXAFS simulations by about 0.3 Å. The FT of EXAFS spectra did not change strongly due to cycling. Minor changes were observed in the $(Co_{0.7}Mn_{0.3})O_x$ spectra before and after cycling, nevertheless, these changes are not prominent, thus not conclusive.

The two prominent peaks were simulated in the Co-K edge: the metal-oxygen distance at 1.87 Å, which is a typical distance for octahedral $Co^{3+}O_6$ cations,[79] and the metal-metal distance around 2.81 Å, associated with metal-metal di-µ-oxo bridge.[48,80] No clear peaks are observed at a higher reduced distance, suggesting a lack of long-range order in the films.

On the other hand, the same peaks were observed in the Mn-K edge, with similar interatomic distances. The peak at 1.87 Å suggests the presence of octahedral $Mn^{3+/4+}O_6$ cations[81,82] and Mn-Mn di-µ-oxo bridge[83] is confirmed by the peak positioned at 2.81 Å. Moreover, an extra Mn-O distance of about 2.30 Å was included in the simulations, improving the fit significantly. This structural motif has been associated with $Mn^{3+}$-O with a Jahn-Teller elongation or $Mn^{2+}$-O.[84] A distance around 2.3 Å has been typically observed in $Mn^{2+}$O in spinel-type oxides.[85] As in the Co-K edge, no clear peaks of additional M-M scatters were observed at higher reduced distance.

EXAFS of the Mn-K and Co-K edge indicated that the pristine films were electrodeposited as a layered hydroxide and did not suffer significant changes in the local structure due to cycling. The decrease in the Mn-M interatomic distance in $(Co_{0.7}Mn_{0.3})O_x$ (Table 2) in comparison to the Mn-Mn distance previously reported for the electrodeposited $MnO_x$[47] suggests that Mn and Co are in the same phase. A mixed Co,Mn-containing phase agrees with the well distributed Mn and Co content on the surface (Figure 1b). Nonetheless, the presence of other minor Mn- or Co-phases cannot be rigorously discarded.

The Co-K and Mn-K edge XANES spectra were used to analyze the nominal metal oxidation state by the calibration of the edge energy with references (Figure S10 and Table S5); $Co^{2+}O$, $Co^{2.6+}_3O_4$ and $LiCo^{3+}O_2$ were the references for Co-K edge, and $Mn^{2+}O$, $Mn^{2.6+}_3O_4$, $Mn^{3+}_2O_3$ and $Mn^{4+}O_2$ for Mn-K edge. The average bulk Co oxidation state was between 2.7+ and 2.8+ in all the samples, indicating that any redox changes that may have occurred to Co due to potential cycling did not influence the chemical state of the bulk. Co oxidation states near 3+ were previously found beneficial for catalytic activity in layered Co oxides.[50]

In the case of Mn-K edge in the $(Co_{0.7}Mn_{0.3})O_x$ films, the averaged bulk Mn oxidation state was 3.7+ and did not change after cycling. However, the averaged Mn oxidation state of $(Co_{0.7}Mn_{0.3})O_x$ is 0.2 oxidation states higher than the previously studied $MnO_x$ films (black line in Figure 5d).[47] Phases containing octahedral $Mn^{3+}$ and $Mn^{4+}$ sites have been found as active materials,[86–92] yet the predominance of $Mn^{4+}$ over $Mn^{3+}$ has a negative impact by making the material inactive or less active,[93] thus increasing the overpotential.[20] In spinel-type catalysts, octahedral $Mn^{3+}$ (rather than tetrahedral $Mn^{2+}$) has been proposed as the active site,[94] which is predicted by the model proposed by Suntivich et al.[95], where occupancy of the $e_g$ orbitals close to one ($Mn^{3+}$) shows the lowest overpotential. Lastly, the tetrahedral $Mn^{2+}$ has been reported as unlikely relevant for OER (only playing a role in the active phase formation).[96]

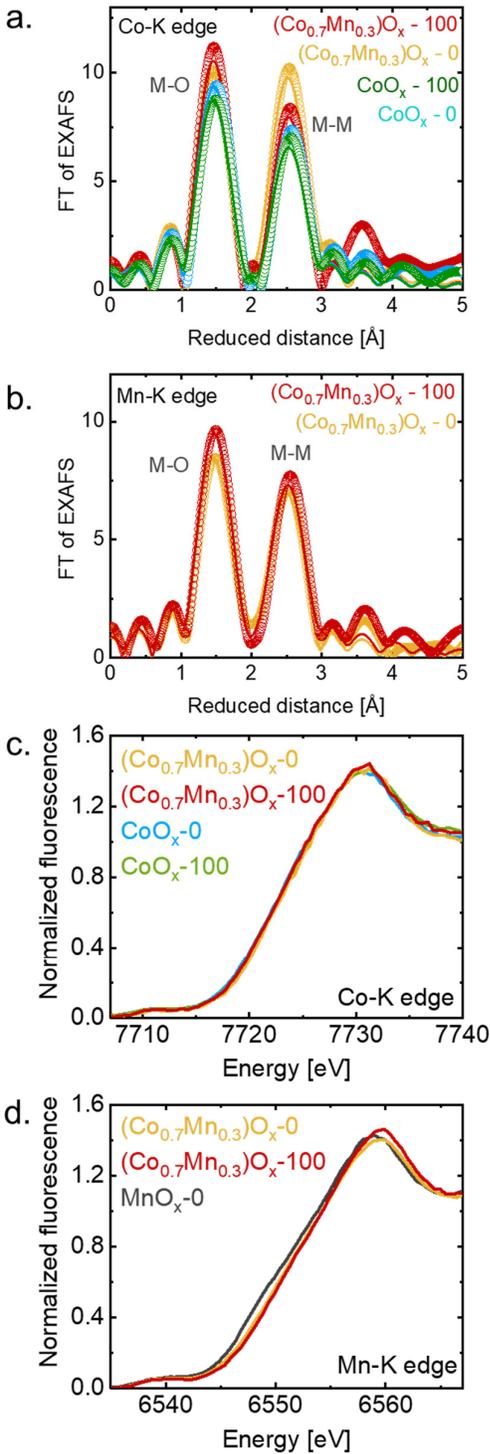

**Figure 5.** FT of EXAFS spectra of: **a.** Co-K edge and **b.** Mn-K edge, collected on pristine $CoO_x$ and $(Co_{0.7}Mn_{0.3})O_x$, and after 100 cycles. The open symbols represent the experimental spectra and the solid lines represent the simulations. The reduced distance is shorter than the precise distance obtained by EXAFS simulations by about 0.3 Å, **c.** XANES spectra of Co-K edge and **d.** Mn-K edge. The Mn-K edge collected on $MnO_x$ (black line) from a previous report was added for comparison.[47] Dataset in ref.[49]

**Table 1.** EXAFS absorber-scatter averaged distance (R), neighbouring atoms number (N) and Debye-Waller factor (σ) as determined by simulation of the $k^3$-weighted EXAFS spectra at the Co-K edge for pristine $CoO_x$ ($CoO_x$-0), $CoO_x$ after 100 cycles ($CoO_x$-100), pristine $(Co_{0.7}Mn_{0.3})O_x$ ($(Co_{0.7}Mn_{0.3})O_x$-0) and $(Co_{0.7}Mn_{0.3})O_x$ after 100 cycles ($(Co_{0.7}Mn_{0.3})O_x$-100). Shells were simulated using phase functions from previously reported structures.[97,98]

| Sample | Parameter | Co–O1 | Co–M[b] | R – factor |
|---|---|---|---|---|
| $CoO_x$-0 | N | 5.7 | 3.16 | |
| | R (Å) | 1.88 | 2.81 | 2.83 % |
| | σ (Å) | 0.05[a] | 0.05[a] | |
| $CoO_x$-100 | N | 5.3 | 3.0 | |
| | R (Å) | 1.87 | 2.81 | 2.50 % |
| | σ (Å) | 0.05[a] | 0.05[a] | |
| $(Co_{0.7}Mn_{0.3})O_x$-0 | N | 6.0 | 4.6 | |
| | R (Å) | 1.87 | 2.79 | 1.47 % |
| | σ (Å) | 0.05[a] | 0.05[a] | |
| $(Co_{0.7}Mn_{0.3})O_x$-100 | N | 6[a] | 3.5 | |
| | R (Å) | 1.87 | 2.79 | 3.18 % |
| | σ (Å) | 0.05[a] | 0.05[a] | |

[a] indicates fixed values (not simulated). [b] M indicates Mn or Co.

**Table 2.** EXAFS absorber-scatter averaged distance (R), neighboring atoms number (N) and Debye-Waller factor (σ) as determined by simulation of the $k^3$-weighted EXAFS spectra at the Mn-K edge for pristine $(Co_{0.7}Mn_{0.3})O_x$ ($(Co_{0.7}Mn_{0.3})O_x$ -0) and $(Co_{0.7}Mn_{0.3})O_x$ after 100 cycles ($(Co_{0.7}Mn_{0.3})O_x$ -100). Shells were simulated using phase functions from a previously reported structure.[81]

| Sample | Parameter | Mn–O1 | Mn–O2 | Mn–M[b] | R – factor |
|---|---|---|---|---|---|
| $(Co_{0.7}Mn_{0.3})O_x$-0 | N | 5.1 | 1[a] | 3.4 | |
| | R (Å) | 1.87 | 2.36 | 2.82 | 0.73 % |
| | σ (Å) | 0.05[a] | 0.05[a] | 0.05[a] | |
| $(Co_{0.7}Mn_{0.3})O_x$-100 | N | 5[a] | 1[a] | 3.7 | |
| | R (Å) | 1.87 | 2.31 | 2.83 | 1.15 % |
| | σ (Å) | 0.05[a] | 0.05[a] | 0.05[a] | |

[a] indicates fixed values (not simulated). [b] M indicates Mn or Co.

The metal-K edges previously discussed can identify bulk material changes, but they might neglect changes occurring only at the near-surface region. As catalysis is a surface process, the films were also analyzed using the total electron yield (TEY) of the Co-$L_3$ and Mn-$L_3$ edges, whose escape depth is of a few nm (2.6±0.3 nm for a similar oxide at the Mn-L edge).[99] The Co-$L_3$ spectrum of the pristine $CoO_x$ showed clear features of the $Co^{2+}$ references (highlighted in blue in Figure 6a), indicating the dominant $Co^{2+}$ content, which differs from the Co-K edge spectrum. Yet, after 100 cycles the spectrum changed drastically, and the $Co^{2+}$ features were no longer present. After 100 cycles, only one prominent peak is observed, which closely resembles the spectrum of the $Co^{3+}$ reference, $LiCoO_2$ (highlighted in orange in Figure 6a). An apparent oxidation of the near-surface region was observed after 100 cycles. The increase in the oxidation state can be attributed to the oxidation of $Co^{2+}$ sites to $Co^{3+}$ sites, likely at potential values around 1.42 V vs. RHE.[50,72] A redox peak was barely observed in the CV experiments of $CoO_x$ (inset in Figure 2a) at around 1.5 V vs. RHE, which can be assigned to the oxidation of a small number of $Co^{2+}$ sites. Such redox peak is not observed in $(Co_{0.7}Mn_{0.3})O_x$ studied in the same range of potential. The unlikely relevancy of $Co^{2+}$ sites in OER suggests that the conversion of $Co^{2+}$ into $Co^{3+}$ sites should benefit the catalytic activity,[50,72] yet the catalytic activity did not increase but rather decreased according to RRDE measurements.

Another possibility, that should not be discarded, is that the catalytically less relevant $Co^{2+}$ ions are lost from the near surface region since it is well soluble in aqueous solutions.[100] These ions could come either from minor $Co^{2+}$ phases or from the $Co^{2+}$-rich electrodeposition electrolyte. The latter is less likely as the samples were soaked in DI water to remove the electrodeposition electrolyte. However, $CoO_x$ is not stable at pH 7 at OCP and the formation of $Co^{2+}$ is thus expected due to the cleaning procedure.[101] In contrast to $CoO_x$, the Co-$L_3$ edge spectra of $(Co_{0.7}Mn_{0.3})O_x$ did not significantly change after cycling and resemble the $Co^{3+}$ reference ($LiCoO_2$), which indicates no major changes at near-surface region of the catalyst after 100 cycles.

The Mn-$L_2$ edge of $(Co_{0.7}Mn_{0.3})O_x$ showed two prominent peaks (highlighted in green in Figure 6b) that resemble the $MnO_2$ (and partially $Mn_2O_3$) reference and no evident changes are observed due to cycling. The near-surface region exhibits an oxidation state between 3+ and 4+, which agrees with Mn-K edge measurements, where an oxidation state of the bulk of the material was estimated to be 3.7+. In comparison to the previously reported $MnO_x$ films, the averaged Mn oxidation state was 0.2 lower than the herein studied $(Co_{0.7}Mn_{0.3})O_x$ films (Table S5), yet in both cases the bulk averaged Mn oxidation state remained unaffected after 100 cycles. On the other hand, the near-surface region of the previously reported $MnO_x$ suffered an oxidation towards $Mn^{4+}$,[47] which affected the catalytic activity by decreasing the current over cycling. The Mn oxidation was identified as an irreversible change; therefore, the catalytic current did not fully recover after the OCP break. Such effect is not observed in the $(Co_{0.7}Mn_{0.3})O_x$ films since the near-surface region (as well as the bulk) remained unaffected also at the Mn-$L_3$ edge. These observations indicate that Mn is more stable in a slightly higher oxidation state (3.7+), which was initially promoted by to the presence of Co in $(Co_{0.7}Mn_{0.3})O_x$. Since $Mn^{4+}$ and $Mn^{2+}$ tend to comproportionate to $Mn^{3+}$ in alkaline media,[25] the formation of Mn-based oxide with a slightly more oxidized Mn could promote more stable Mn sites. Mn oxides containing $Mn^{3+}$ and $Mn^{4+}$ have shown more stability and lower overpotential.[89,90,93,102,103]

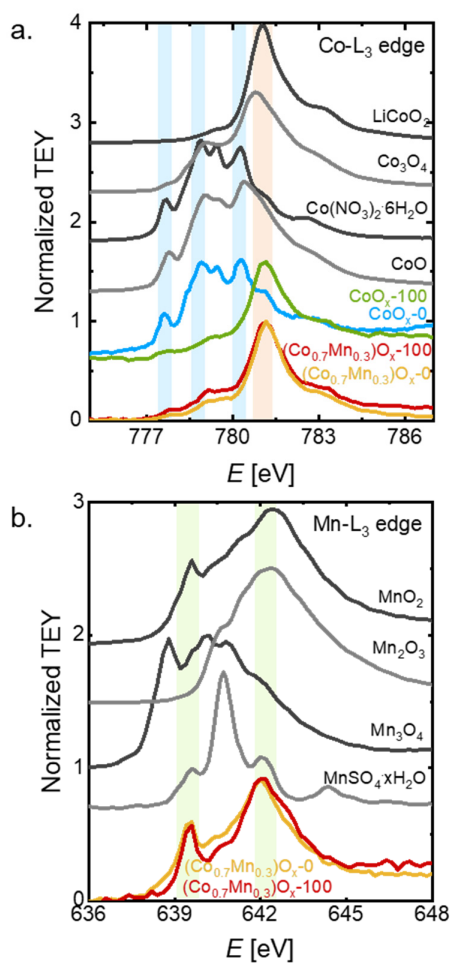

**Figure 6. a**. XAS spectra of: **a.** Co-$L_3$ edge and **b.** Mn-$L_3$ edge, collected on pristine $CoO_x$ and $(Co_{0.7}Mn_{0.3})O_x$, and after 100 cycles. The light-colored regions are added to help assign the relevant peaks in Co-$L_3$ edge (orange highlights CoOx-100, (Co,Mn)Ox-0 and (Co,Mn)Ox-100 peaks; blue highlights CoOx-0 peaks, and green highlights (Co,Mn)Ox-0 and (Co,Mn)Ox-100 peaks). $MnO_2$, $Mn_2O_3$, $Mn_3O_4$, $Mn(SO_4)_2\cdot xH_2O$, CoO, $Co(NO_3)_2\cdot 6H_2O$, $Co_3O_4$ and $LiCoO_2$ were used as references. Dataset in ref.[49]

In summary, using the Co-K and Mn-K edge the films were characterized as layered hydroxides, whose local structure and bulk averaged oxidation state did not significantly change due to continuous potential cycling. Moreover, the analysis of the Co-$L_3$ edge suggested the oxidation of the near-surface region of the $CoO_x$ films after 100 cycles, whereas $(Co_{0.7}Mn_{0.3})O_x$ did not suffer significant changes.

The electronic structure of the $CoO_x$ film seems to be significantly affected by the incorporation of about 30 % of Mn oxide into structure, which had a beneficial effect in the stabilization under OER conditions. Although, it is not trivial to find a single explanation for this effect, herein we address the most likely explanations.

The introduction of a second metallic site, in this case Mn, generates more oxygen vacancies,[104–106] which improves the redox activity of the material.[66] Such effect may prevent the high coverage (or even saturation) of the catalyst with unreacted intermediates or oxygen. Co remained in an oxidation state close to 3+ being optimal for the OER,[50,95] while Mn in $(Co_{0.7}Mn_{0.3})O_x$ was in oxidation state 3.7+ independent of potential cycling. The latter is different for $MnO_x$, where we previously argued that Mn oxidation is the main irreversible cause of activity loss.[47] Thus, it is plausible that both Co and Mn remain active in $(Co_{0.7}Mn_{0.3})O_x$. Moreover, even though Co oxides are considered promising OER catalysts, they do not own sufficiently high electrical conductivity,[107–109] which is a desirable feature in OER catalysts[110–112] to favor the rate of electron transport through the material.[111] The introduction of Mn oxide into the Co oxide structure could improve the vacancies variation among the different metallic sites,[112] but also because the incorporation of Mn into Co oxides optimizes the electrical conductivity.[45]

## Conclusion

Na-containing layered $CoO_x$ and $(Co_{0.7}Mn_{0.3})O_x$ films were electrodeposited in 0.1 M NaOH solution, using a complexing agent for the stabilization of the ions. Incorporating $MnO_x$ into the $CoO_x$ does not significantly change the onset of OER and results in a slight decrease in the overpotential at 10 mA cm$^{-2}$, $\eta_{10}$, in the 2$^{nd}$ cycle. After 100 cycles, $\eta_{10}$ of $CoO_x$ and $(Co_{0.7}Mn_{0.3})O_x$ are identical within error with a slightly lower average of the latter. Moreover, The Tafel slope of $CoO_x$ tended to increase over cycling, whereas the Tafel slope of $(Co_{0.7}Mn_{0.3})O_x$ was mostly constant during 100 cycles, indicating that $CoO_x$ may not efficiently support high currents for long durations. Often, there is a trade-off between catalytic activity and stability.[6] While we showed that 30 % Mn in layered $CoO_x$ only had a minor effect on activity, it stabilized the activity during potential cycling under OER conditions.

As expected from the electrocatalytic trends, no changes were identified by XAS in $(Co_{0.7}Mn_{0.3})O_x$. We discussed that the absence of valence changes can explain the electrocatalytic trends. Yet, the addition of Mn to $CoO_x$ may have further beneficial effects such as the formation of more oxygen vacancies in the catalyst, which might avoid the saturation of the surface with unreacted intermediates or products and a likely increase in electrical conductivity. Overall, the addition of 30% Mn to $CoO_x$ successfully stabilized the catalyst material and its activity during OER cycling. The development of new strategies for the stabilization of catalyst materials and insight into the origin of stabilization are essential for the future knowledge-guided design of durable electrocatalysts for electrolyzers.

## Experimental Section

### Materials

$Co(NO_3)_2 \cdot 6H_2O$ (≥ 99.999 %), $Co_3O_4$ (99.99), CoO (99.99 %), $LiCoO_2$ (>99.8 %), $Mn(NO_3)_2 \cdot 4H_2O$ (≥ 99.99 %), $Mn(SO_4)_2 \cdot xH_2O$ (99.99), $MnO_2$ (≥ 99 %), $Mn_3O_4$ (≥ 97 %), $Mn_2O_3$ (≥ 99.9 %), L-(+)-Tartaric acid (≥ 99.5 %) and (2 M and 0.1 M) NaOH solutions were ordered from Sigma-Aldrich. Graphite foil (≥ 99.8 %) with a thickness of 0.254 mm ordered from VWR. All reactants were used as received, without any further treatment. Solutions were prepared with deionized water (>18 MΩ cm).

### Films electrodeposition

**$CoO_x$ films:** 0.6 mmol of $Co(NO_3)_2 \cdot 6H_2O$ and 6 mmol of L-(+)-tartaric acid were dissolved in a small volume of deionized water (approx. 1 mL). 120 mL of Ar-purged 2 M NaOH solution were added slowly to the previous solution while stirring, changing from colorless to beige.

**$(Co_{0.7}Mn_{0.3})O_x$ films:** were prepared with a similar procedure to $CoO_x$ using a mixture of 0.6 mmol of $Co(NO_3)_2 \cdot 6H_2O$ and 0.6 mmol of $Mn(NO_3)_2 \cdot 4H_2O$ as precursor solution. All other parameters remained the same.

The electrodeposition of the films was performed in a three-electrode cell made from a three-neck round-bottom flask and using a Gamry Reference 600+ potentiostat. The distance between the necks and thus the electrodes was kept lower than 1 cm. The working electrodes were either a glassy carbon disk (4 mm diameter; HTW Sigradur G) in a rotating disk electrode (RDE) or graphite paper (Alfa Aesar). The unrotated RDE was mounted onto a commercial rotator (ALS RRDE-3A Ver 2.0). We used a saturated calomel reference electrode (SCE; ALS RE-2BP) and a graphite rod (redox.me, HP-III, High Pure Graphite) as the counter electrode. The galvanostatic deposition was performed at 150 µA cm$^{-2}$ until a charge density of 40 mC cm$^{-2}$ was reached.

### Electrochemical measurements

The detailed protocol for electrocatalytic investigations is documented in Table S1 for glassy carbon electrodes and in Table S2 for graphite foil. The measurements on glassy carbon electrodes were carried out using two Gamry Reference 600+ potentiostats connected as a bipotentiostat in a single-compartment three-electrode electrochemical cell made of polymethyl pentene (ALS) filled with about 60 mL solution of 0.1 M NaOH. A commercial rotator (ALS RRDE3-A Ver 2.0) was used with commercial rotating ring-disk electrodes (RRDE) with exchangeable disks of 4 mm diameter and a Pt ring with inner ring diameter of 5 mm and outer diameter of 7 mm. The graphite foil was clamped in the same cell as the RRDE. A coiled platinum wire was used as a counter electrode and a SCE (ALS RE-2BP) as a reference electrode, which was calibrated daily against a commercial reversible hydrogen electrode (RHE; Gaskatel HydroFlex). The electrochemical experiments were performed at constant controlled temperature of 25.0 °C. The ring was set to detect oxygen at 0.4 V vs. RHE as calibrated previously.[55] Before any experiment, the electrolyte was purged with Ar for at least 30 minutes. The ohmic drop (also called iR$_u$ drop) was corrected during post-processing by subtraction of iR$_u$ from the measured potentials, where i and R$_u$ are the measured current and uncompensated resistance, respectively. All potentials are given relative to the reversible hydrogen electrode (RHE).

The Tafel slope was also calculated with a fitting of potential as function of the logarithm of the current, using the cathodic half-cycle of the cyclic voltammetry of iR$_u$-corrected data in the range between 1.71 and 1.76 V vs. RHE. The electrodes were swept at 100 mV s$^{-1}$ and rotated at 1600 rpm. The Tafel slope was obtained by linear regression of the iR$_u$-corrected potential (E-iR$_u$) against log$_{10}$(i). The error represents the standard deviation of three independently prepared electrodes.

### Scanning electron microscopy (SEM) and energy dispersive X-ray spectroscopy (EDX)

The morphology of the samples was studied using a Zeiss LEO Gemini 1530 scanning electron microscope, with an acceleration voltage of 3 keV in high vacuum (approximately $10^{-9}$ bar) and using a secondary electron inLens detector. Images were taken in several regions of the sample to get representative data. EDX measurements were performed using a Thermo Fischer detector with an acceleration voltage of 12 keV.

*X-ray absorption spectroscopy (XAS)*

All XAS data were collected at an averaged nominal ring current of 300 mA in top-up and multi-bunch mode at the BESSY II synchrotron operated by Helmholtz-Zentrum Berlin.

Soft XAS measurements at the Mn-L edges were conducted using the LiXEdrom experimental station at the UE56/2 PGM-2 or U49-2 PGM-1 beamline.[113] Reference samples were measured as finely dispersed powders attached to carbon tape and electrodeposited samples were measured on graphite foil (Alfa Aesar). All samples were measured at room temperature and in total electron yield (TEY) mode and with horizontally linear polarization of the beam. The TEY measurements were carried out by collecting the drain current from the sample. The sample holder was connected to an ammeter (Keithley 6514). In order to avoid radiation damage, the incoming photon flux was adjusted to get a TEY current from the sample of around 10 pA. In addition, the sample was kept as thin as possible. XAS spectra for each sample were collected at a few locations to ensure representativity of the data and further minimize radiation damage and local heating. The energy axis was calibrated using a Mn-L edge spectrum of $MnSO_4$ as a standard where the maximum of the $L_3$-edge was calibrated to 641 eV. This reference was calibrated against molecular oxygen as described elsewhere.[114,115] All spectra were normalized by the subtraction of a straight line obtained by fitting the data before the $L_3$ edge and division by a polynomial function obtained by fitting the data after the $L_3$ edge.

Hard XAS measurements were performed at the KMC-2 or KMC-3 beamlines.[116,117] Co-K and Mn-K edge references were collected at KMC-3. Samples at Co-K edge and Mn-K edge as well as a few references were collected at KMC-2. Two refences spectra were compared to confirm the correct energy calibration. At KMC-3, spectra were recorded in fluorescence mode using a 13-element silicon drift detector (SDD) from RaySpec. The used monochromator was a double-crystal Si (111), and the polarization of the beam was horizontal. Reference samples were prepared by dispersing a thin and homogeneous layer of the ground powder on Kapton tape. After removing the excess material, the tape was sealed, and the excess of Kapton was folded several times to get 1 cm × 1 cm windows. The energy was calibrated using a Co metal foil (fitted reference energy of 7709 eV in the first derivative spectrum) with an accuracy ±0.1 eV. Up to three scans of each sample were collected to $k$ = 14 Å$^{-1}$.

At KMC-2, the general used setup was organized as it follows: $I_0$ ionization chamber, sample, $I_1$ ionization chamber or FY detector, energy reference and $I_2$ ionization chamber. The used double monochromator consisted of two Ge-graded Si(111) crystal substrates[118] and the polarization of the beam was linear horizontal. Reference samples were prepared by dispersing a thin and homogenous layer of the powder on Kapton tape, after removing excess of powder, the tape was folded several times to get 2 cm x 1 cm windows. Reference samples were measured in transmission mode between two ion chambers detector at room temperature. Electrodeposited samples were measured on graphite foil in fluorescence mode with a Bruker X-Flash 6/60 detector. Energy calibration of the X-ray near edge structure (XANES) was made with the corresponding metal foil, setting the inflection point for Mn at 6539 eV. All spectra were normalized by the subtraction of a straight line obtained by fitting the data before the K edge and division by a polynomial function obtained by fitting the data after the K edge. The Fourier transform (FT) of the extended X-ray absorption fine structure (EXAFS) was calculated between 40 and 440 eV (3.2 to 10.7 Å$^{-1}$) above the K edge ($E_0$ = 6539 eV for Mn and $E_0$ = 7709 eV for Co). A cosine window covering 10 % on the left side and 10 % on the right side of the EXAFS spectra was used to suppress the side lobes in the FTs. EXAFS simulations were performed using the software SimXLite. After calculation of the phase functions with the FEFF8-Lite [119] program (version 8.5.3, self-consistent field option activated). Atomic coordinates of the FEFF input files were generated from Mn- and Co-based oxide structures,[81,97,98] the EXAFS phase functions did not depend strongly on the details of the used model. An amplitude reduction factor (S0$^2$) of 0.7 was used. The EXAFS simulations were optimized by the minimization of the error sum obtained by summation of the squared deviations between measured and simulated values (least-squares fit). The fit was performed using the Levenberg–Marquardt method with numerical derivatives.

The authors do not declare a conflict of interest

The data that support the findings of this study are openly available on Figshare at DOI 10.6084/m9.figshare.18415520.

## Acknowledgements

The authors thank Dr. Max Baumung, Joaquín Morales-Santelices and Sepideh Madadkhani for helping in data collection. Frederik Stender is acknowledged for writing the electrochemistry analysis script and Dr. Petko Chernev for permission to use his software SimXLite. Dr. Michael Haumann and Dr. Ivo Zizak are thanked for support at the beamline station. We thank Helmholtz-Zentrum Berlin (HZB) for the allocation of synchrotron radiation beamtime and acknowledge the HZB CoreLab CCMS (Correlative Microscopy and Spectroscopy) for training and advising in SEM. This project has received funding from the European Research Council (ERC) under the European Union's Horizon 2020 research and innovation programme under grant agreement No 804092.

**Keywords:** catalyst stability • bimetallic oxides • catalyst activation • oxygen evolution reaction • Co-based oxides

# Stabilization of Co oxide during cyclic voltammetry by the introduction of Mn oxide in alkaline media

Javier Villalobos,[a] Dulce M. Morales,[a] Denis Antipin,[a] Götz Schuck, [b] Ronny Golnak,[c] Jie Xiao,[c] Marcel Risch[a]*

[a] Nachwuchsgruppe Gestaltung des Sauerstoffentwicklungsmechanismus
Helmholtz–Zentrum Berlin für Materialien und Energie GmbH
Hahn-Meitner Platz 1, Berlin 14109, Germany
[b] Abteilung Struktur und Dynamik von Energiematerialien
Helmholtz–Zentrum Berlin für Materialien und Energie GmbH
Hahn-Meitner Platz 1, Berlin 14109, Germany
[c] Department of Highly Sensitive X-ray Spectroscopy
Helmholtz–Zentrum Berlin für Materialien und Energie GmbH
Albert-Einstein-Straße 15, Berlin 12489, Germany

Corresponding author
marcel.risch@helmholtz-berlin.de

10 pages, 5 supplementary tables, 10 supplementary figures

**Table S1.** General protocol for electrochemical data collection on an RRDE station with samples on glassy carbon disks. All potentials are reported vs. RHE. Electrolyte was 0.1 M NaOH. Blanket indicates no purge.

| Step | Conditions |
|---|---|
| **1.** Cleaning | Clean and polish electrodes, cells and any other tool properly. |
| **2.** Calibration of reference electrodes | OCP against commercial RHE electrode |
| **3.** Argon purge at OCP | At least 30 minutes |
| **4.a.** Ring EIS | Frequency: 1 MHz – 1 Hz. Points/decade: 10. OCP and take note of $R_u$ |
| **4.b.** Disk EIS | Frequency: 1 MHz – 1 Hz. Points/decade: 10. OCP and take note of $R_u$ |
| **5.** Disk CV: ECSA* | Hold 10 s at 1.0 V. Potential window: 0.95 V – 1.05 V. Scan rates: 50, 100, 150, 200, 250, 300, 350, 400, 450, 500 mV s$^{-1}$. Cycles: 3. Rotation: 0 rpm. Purge: Blanket. No dynamic $iR_u$ compensation. |
| **6.a.** Ring conditioning | Hold ring potential for 1800 s at 0.40 V. |
| **6.b.** Ring: CA ($O_2$ detection) | Hold ring potential at 0.40 V. |
| **6.c.** Disk CV: OER | Potential window: 1.40 V - 1.80 V. Scan rate: 100 mV s$^{-1}$. Step size: 2 mV. Cycles: 100. Rotation: 1600 rpm. Purge: yes. No dynamic $iR_u$ compensation. |
| **7.** Disk CV: ECSA* | Hold 10 s at 1.0 V. Potential window: 0.95 V – 1.05 V. Scan rates: 50, 100, 150, 200, 250, 300, 350, 400, 450, 500 mV s$^{-1}$. Cycles: 3. Rotation: 0 rpm. Purge: Blanket. No dynamic $iR_u$ compensation. |
| **8.** Disk OCP | 1800 s |
| **9.a.** Disk EIS | Frequency: 1 MHz – 1 Hz. OCP and take note of $R_u$ |
| **9.b.** Ring EIS | Frequency: 1 MHz – 1 Hz. OCP and take note of $R_u$ |
| **10.a.** Ring conditioning | Hold ring potential for 1800 s at 0.40 V. |
| **10.b.** Ring: CA ($O_2$ detection) | Hold ring potential at 0.40 V. |
| **11.** Disk CV: OER | Potential window: 1.40 V - 1.80 V. Scan rate: 100 mV s$^{-1}$. Step size: 2 mV. Cycles: 10. Rotation: 1600 rpm. Purge: yes. No dynamic $iR_u$ compensation. |

\* not used due to inappropriate data for analysis**.**

**Table S2.** General protocol for electrochemical data collection with samples on graphite foil. All potentials are reported vs. RHE. Electrolyte was 0.1 M NaOH.

| 1. Cleaning | Clean and polish electrodes, cells and any other tool properly. |
|---|---|
| 2. Calibration of reference electrodes | OCP against commercial RHE electrode |
| 3. Argon purge at OCP | At least 30 minutes |
| 4. Foil CV: OER | Potential window: 1.40 V - 1.80 V<br>Scan rate: 100 mV s$^{-1}$<br>Step size: 2 mV<br>Cycles: 100<br>Rotation: 1600 rpm<br>Purge: yes<br>No dynamic iR$_u$ compensation |
| 5. Sample rinsing | Soaked in deionized water for 5 minutes. |

**Table S3.** Average exponent values of the current trends over cycling. The exponents were estimated from Figure 3 at each selected potential and each selected region.

| Samples | Potential | Region's exponent | | | |
|---|---|---|---|---|---|
| | | 1 | 2 | 3 | 1 – OCP[a] |
| CoO$_x$ | 1.55 | -1/25 | -1/25 | -1/25 | -1/10 |
| | 1.64 | -1/25 | -1/25 | -1/25 | -1/30 |
| | 1.70 | -1/25 | -1/25 | -1/25 | -1/10 |
| (Co$_{0.7}$Mn$_{0.3}$)O$_x$ | 1.55 | -1/7 | -1/5 | -1/15 | -1/6 |
| | 1.66 | -1/34 | -1/17 | 1/4 | -1/7 |
| | 1.70 | 1/16 | 1/9 | 1/5 | -1/22 |

[a] region 1 after the OCP break.

**Table S4.** Averaged Tafel slope values for CoO$_x$ and (Co$_{0.7}$Mn$_{0.3}$)O$_x$ at selected cycles. The values are the average of three different samples and standard deviation is reported as error.

| | CoO$_x$ | | (Co$_{0.7}$Mn$_{0.3}$)O$_x$ | |
|---|---|---|---|---|
| Cycle | Average [mV dec$^{-1}$] | Error [mV dec$^{-1}$] | Average [mV dec$^{-1}$] | Error [mV dec$^{-1}$] |
| 2 | 135 | 9 | 86 | 6 |
| 3 | 136 | 9 | 91 | 1 |
| 5 | 136 | 12 | 89 | 3 |
| 10 | 138 | 11 | 88 | 2 |
| 20 | 144 | 15 | 89 | 2 |
| 50 | 149 | 27 | 89 | 3 |
| 100 | 158 | 25 | 88 | 4 |
| 2 | 132 | 9 | 96 | 5 |
| 3 | 133 | 9 | 98 | 4 |
| 5 | 133 | 10 | 99 | 3 |
| 10 | 133 | 11 | 95 | 11 |

**Table S5.** Co and Mn nominal oxidation state of pristine $CoO_x$ and $(Co_{0.7}Mn_{0.3})O_x$, and after 100 cycles. The data was estimated using the metal K edges. The fit equation and graph are shown in Figure S12.

| Sample | Pristine film | After 100 cycles |
|---|---|---|
| | Co nominal oxidation state | |
| $CoO_x$ | 2.70 | 2.72 |
| $(Co_{0.7}Mn_{0.3})O_x$ | 2.77 | 2.70 |
| | Mn nominal oxidation state | |
| $(Co_{0.7}Mn_{0.3})O_x$ | 3.71 | 3.67 |
| $MnO_x$[a] | 3.47 | 3.48 |

[a]obtained from reference.[2]

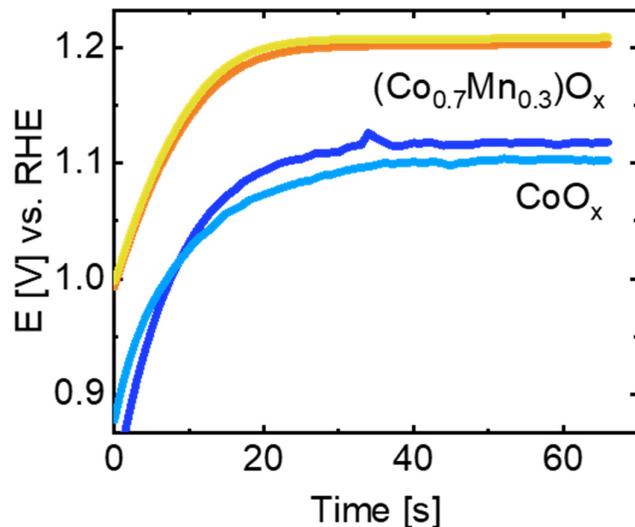

**Figure S1.** Chronopotentiometry during electrodeposition of CoO$_x$ and (Co$_{0.7}$Mn$_{0.3}$)O$_x$ films on graphite foil substrate. Curves are shown in duplicate, corresponding to two independent electrodepositions.

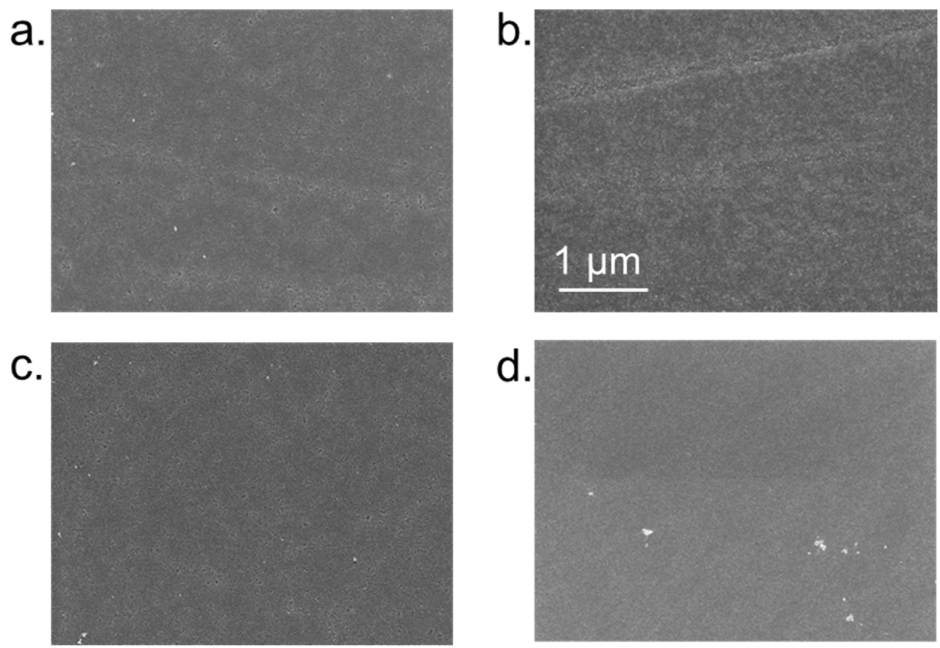

**Figure S2.** SEM images of: a. Pristine CoO$_x$, b. CoO$_x$ after 100 cycles, c. Pristine (Co$_{0.7}$Mn$_{0.3}$)O$_x$ and d. (Co$_{0.7}$Mn$_{0.3}$)O$_x$ after 100 cycles.

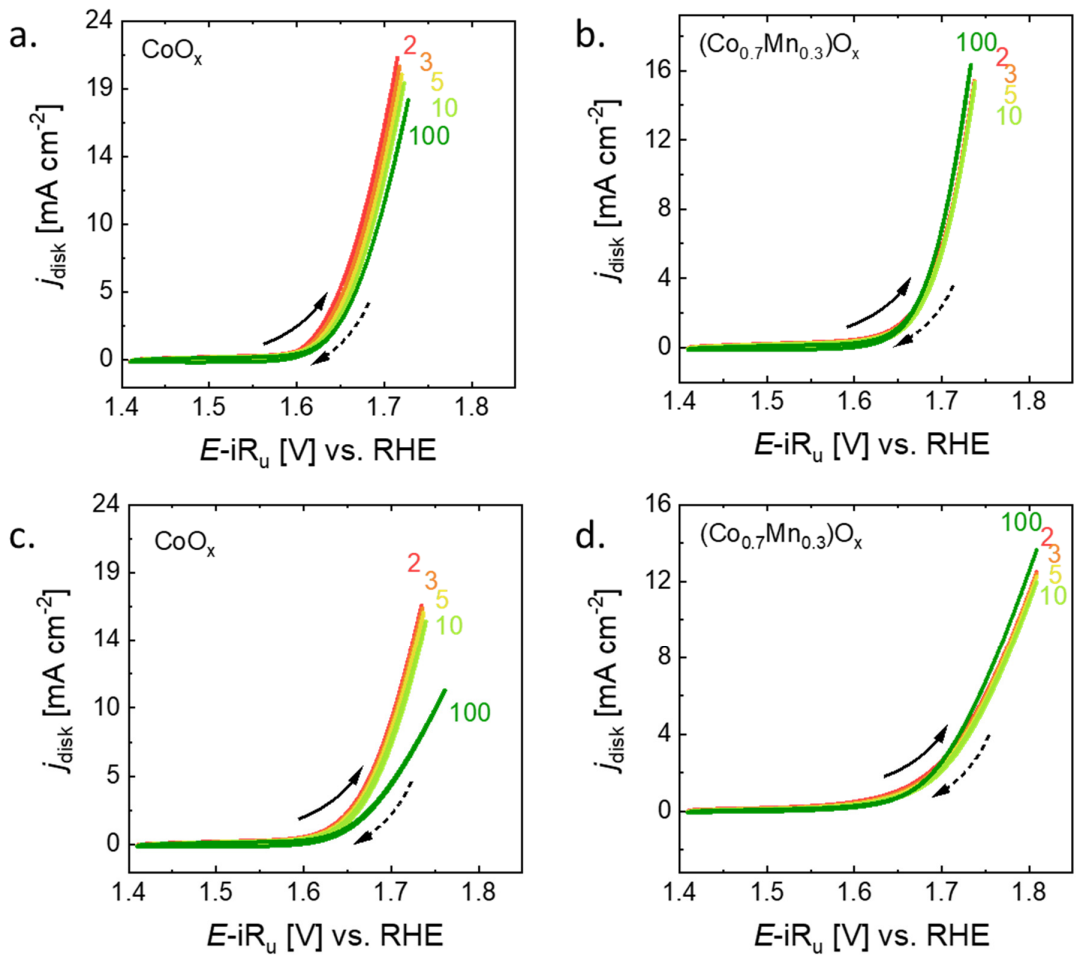

**Figure S3.** Additional CVs of CoO$_x$ (a and c) and (Co$_{0.7}$Mn$_{0.3}$)O$_x$ (b and d). These CVs were collected at the disk with a scan rate of 100 mV s$^{-1}$ in NaOH 0.1 M with a rotation rate of 1600 rpm.

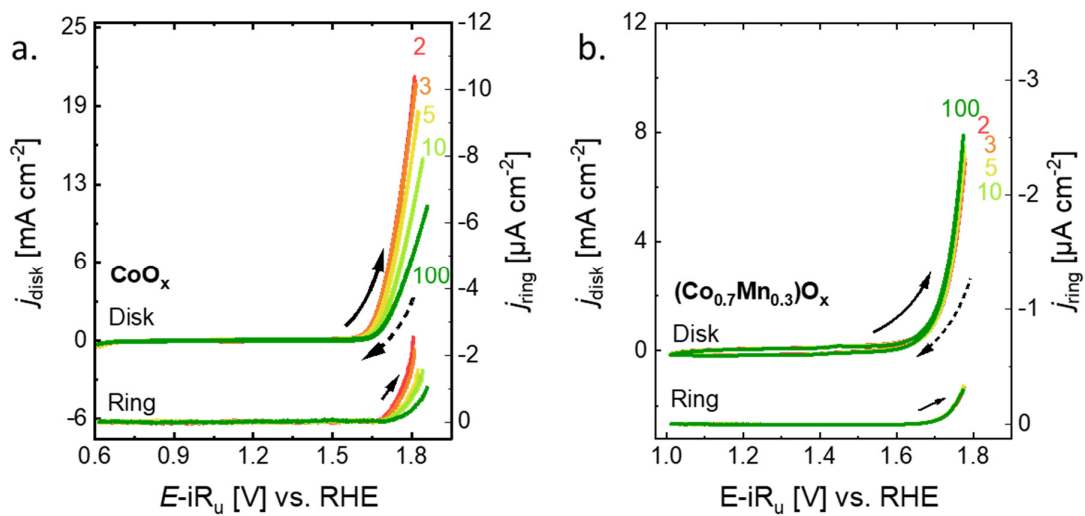

**Figure S4.** Series of CV performed on a: a. CoO$_x$-covered disk and b. (Co$_{0.7}$Mn$_{0.3}$)O$_x$ -covered disk. A constant potential of 0.4 V vs. RHE was applied at the ring. The CV was performed with a scan rate of 100 mV s$^{-1}$ in 0.1 M NaOH with an electrode rotation of 1600 rpm. These samples were measured with lower potential boundary than samples shown in Figure 2. The arrows indicate the direction of the scan.

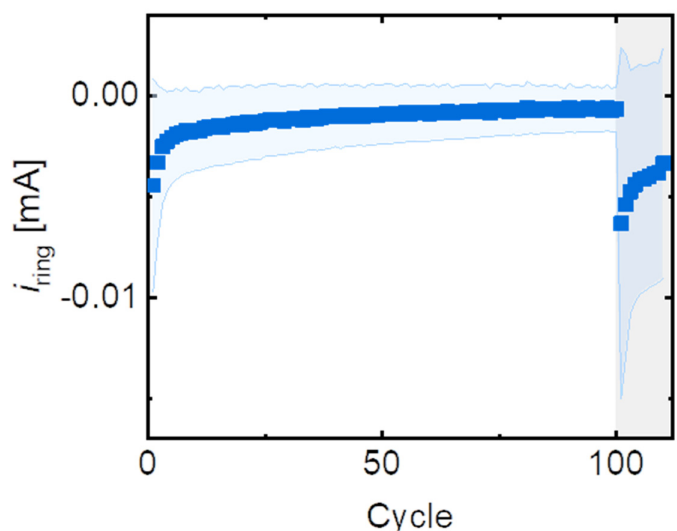

**Figure S5.** Average ring current trend as a function of cycling collected on CoO$_x$.for E-iR$_u$=1.70 V vs. RHE. The blue light-colored area represents the error of three measurements. The grey background represents the 10 additional cycles collected after the OCP break.

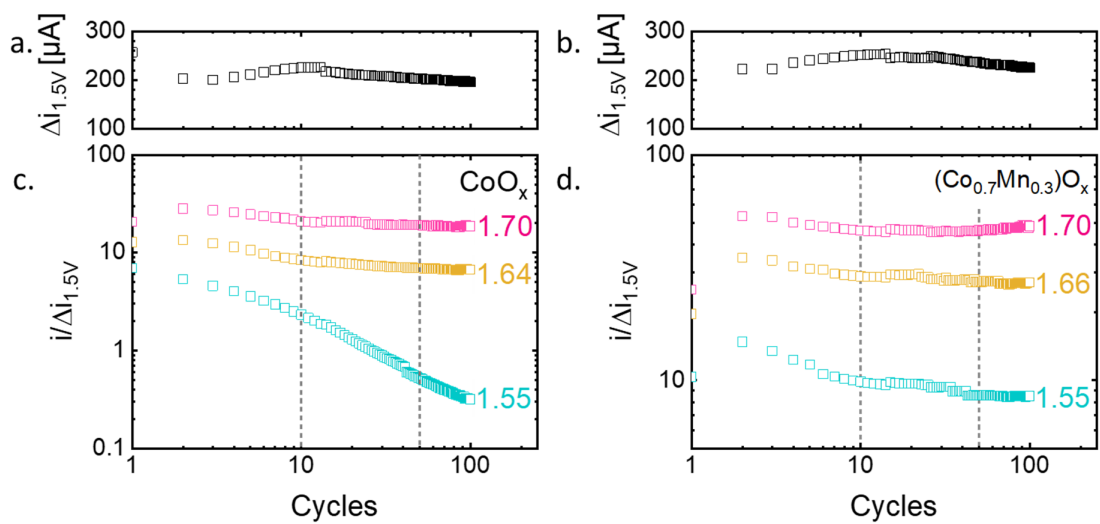

**Figure S6.** $\Delta i_{1.5\,V}$ as function of cycles for the first 100 cycles for a. $CoO_x$ and c. $(Co_{0.7}Mn_{0.3})O_x$ deposited on graphite foil. Current ratio $i/\Delta i_{1.5\,V}$ as a function of cycling at selected potentials for b. $CoO_x$ and d. $(Co_{0.7}Mn_{0.3})O_x$ deposited on graphite foil.

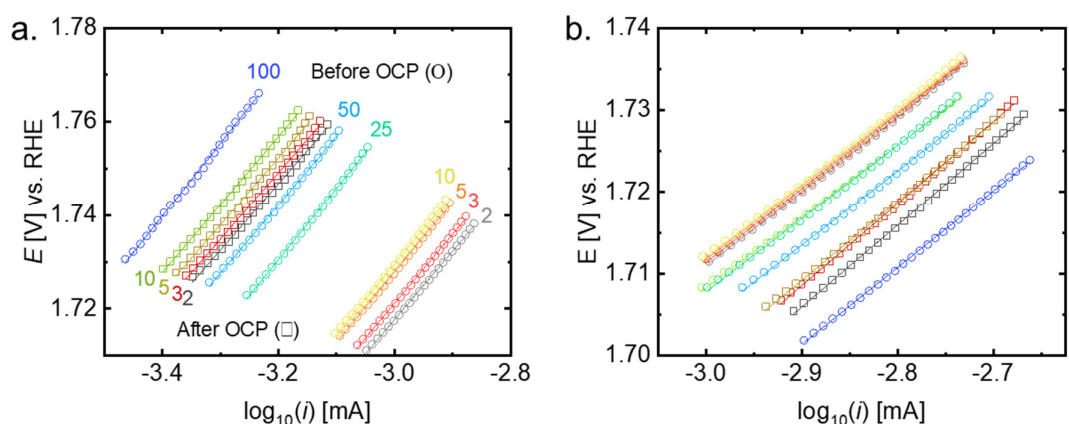

**Figure S7**. Tafel plots of $CoO_x$ (a) and $(Co_{0.7}Mn_{0.3})O_x$ (b) films. Representative plot of Tafel slope calculation for selected cycles (before OCP: 2, 3, 5, 10, 25, 50, 100, and after OCP: 2, 3, 5, 10). The measurements were performed in 0.1 M NaOH. The data was collected with a scan rate 100 mV s$^{-1}$ and the $iR_u$ compensation was done during post-processing. The lines represent the linear fit of $E-iR_u$ as a function of $\log_{10}(i)$, the slope values represent the Tafel slope. Parameters are shown in Table S3.

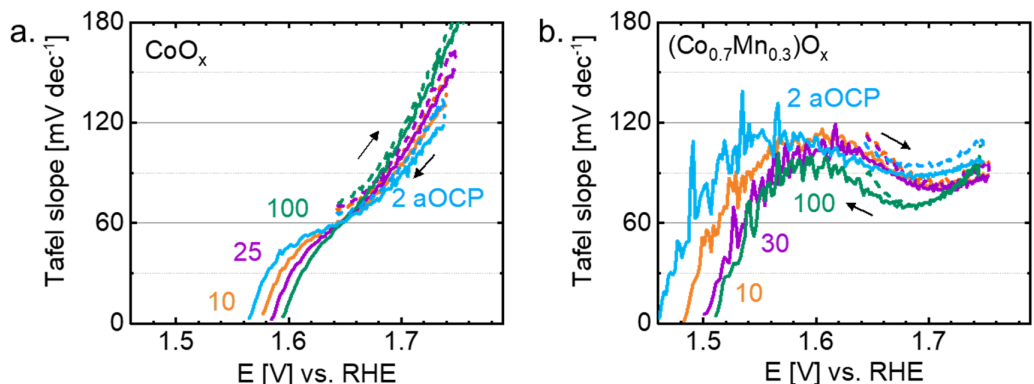

**Figure S8.** Instantaneous Tafel slope as a function of potential estimated for selected cycles (10$^{th}$ cycle, 25$^{th}$ cycle and 100$^{th}$ cycle before the OCP break and 2$^{nd}$ after the OCP break). The data was extracted from the anodic scans in the CVs shown in Figure 2. The instantaneous Tafel slope was calculated by the first derivative of the iR$_u$-corrected potential as function of the logarithm of the current density.

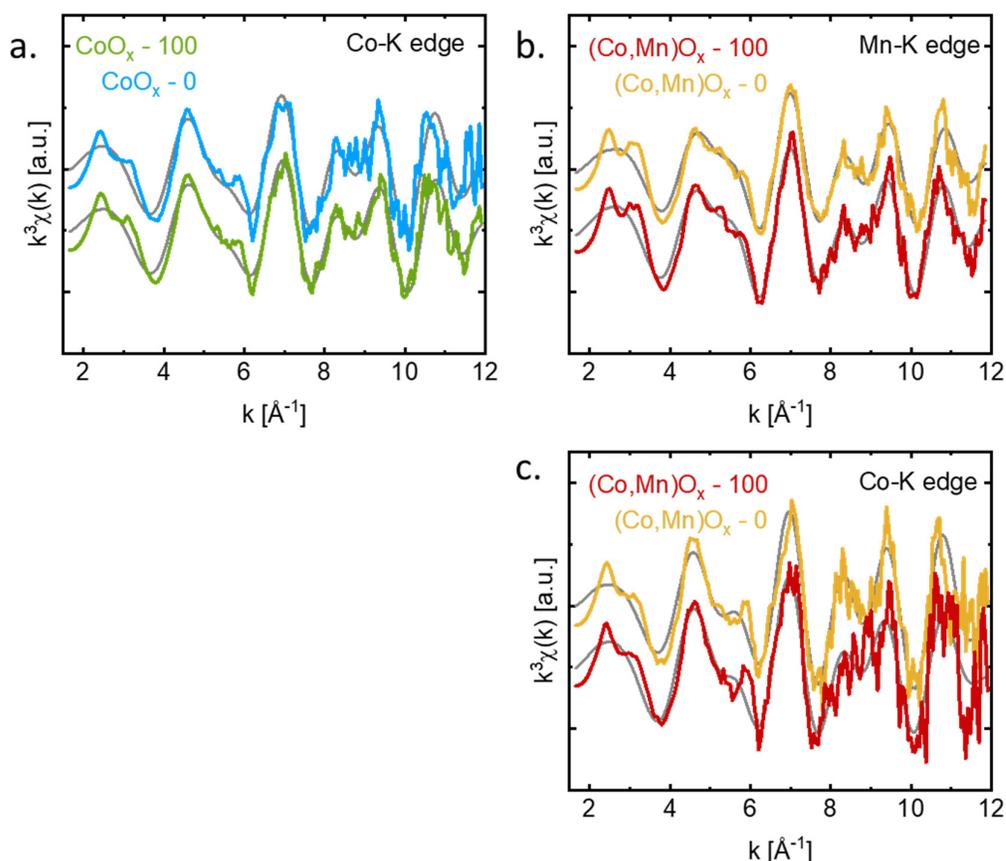

**Figure S9.** k$^3$-weighted EXAFS spectra of pristine CoO$_x$ and (Co$_{0.7}$Mn$_{0.3}$)O$_x$, and after 100 cycles, recorded at the Mn-K edge and Co K-edge. The colored lines represent the measurements and the gray lines the respective EXAFS simulations.

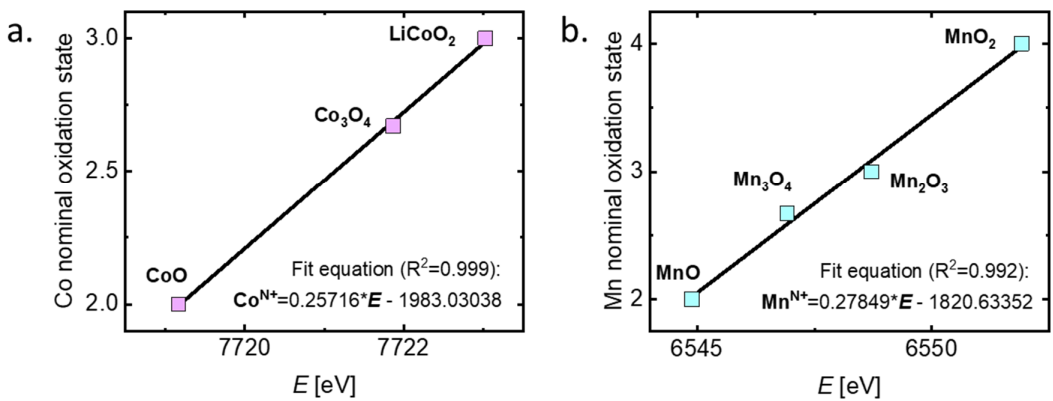

**Figure S10.** Nominal oxidation state of three Co-based references (a) and four Mn-based references (b) as a function of energy of the Co-K edge. The fit equation is shown. CoO, $Co_3O_4$ and $LiCoO_2$ were used as Co references; and MnO, $Mn_3O_4$, $Mn_2O_3$ and $MnO_2$ were used as Mn references. The estimated oxidation states are shown in Table S5. The edge energy was estimated using the integral method ($\mu_1$=1.00 $\mu_2$=0.15).[1]

**References**
1. H. Dau, P. Liebisch and M. Haumann, *Anal. Bioanal. Chem.*, 2003, **376**, 562–583.
2. J. Villalobos, R. Golnak, L. Xi, G. Schuck, M. Risch, *J. Phys. Energy* **2020**, *2*, 034009.